\documentclass[aps,prl,showpacs,twocolumn,superscriptaddress]{revtex4}
\usepackage{graphicx}
\usepackage{bm}
\usepackage{amsmath}

\def\ra{\rangle}
\def\la{\langle}
\def\be{\begin{equation}}
\def\ee{\end{equation}}
\def\bea{\begin{eqnarray}}
\def\eea{\end{eqnarray}}

\def\ua{\uparrow}
\def\da{\downarrow}
\newcommand{\Sin}{{\rm sin}}

\begin{document}
\title{Electron localization/delocalization in incommensurate helical magnets}

\author{Shu Tanaka}
\email[Electronic address: ]{shu-t@spin.phys.s.u-tokyo.ac.jp} 
\affiliation{Department of Physics, The University of Tokyo,
7-3-1, Hongo, Bunkyo-ku, Tokyo 113-0033, Japan}

\author{Hosho Katsura}
\email{katsura@appi.t.u-tokyo.ac.jp}
\affiliation{Department of Applied Physics, The University of Tokyo,
7-3-1, Hongo, Bunkyo-ku, Tokyo 113-8656, Japan}

\author{Naoto Nagaosa}
\email{nagaosa@appi.t.u-tokyo.ac.jp}
\affiliation{Department of Applied Physics, The University of Tokyo,
7-3-1, Hongo, Bunkyo-ku, Tokyo 113-8656, Japan}
\affiliation{Correlated Electron Research Center (CERC),
National Institute of Advanced Industrial Science and Technology (AIST),
Tsukuba Central 4, Tsukuba 305-8562, Japan}
\affiliation{CREST, Japan Science and Technology Agency (JST), Japan}

\begin{abstract}
The electronic states in incommensurate (IC) helical magnets 
are studied theoretically from the viewpoint of the 
localization/delocalization. It is found that in the 
multi-band system with relativistic spin-orbit interaction, 
the electronic wavefunctions show both the extended and localized 
natures along the helical 
axis depending on the orbital, helical wavenumber, and the direction
of the plane on which spins rotate.
Possible realization of this localization is discussed.
\end{abstract}

\maketitle

The helical magnets have been studied for a long term since 
its first discovery by Yoshimori \cite{Yoshimori}. 
Their ground states are determined by the (frustrated) exchange 
interactions and their Fourier transformation $J(q)$.
Various properties including the spin wave excitations are
analyzed theoretically for many materials. 
(See \cite{Nagamiya} for an early review.)
Helical spin structure is now attracting recent revived interests
from the viewpoints of both dielectric and transport properties. 
One example is the ferroelectricity induced by the helical 
magnetic order.
Theoretically, the spin current associated with the noncollinear spin 
configuration is proposed to induce the electric polarization \cite{Katsura}. 
Experimentally, it is now found that this mechanism is at work in 
$R$MnO$_3$ \cite{KimuraNature,Goto,Kenzelmann}
and in other materials \cite{KimuraPRL,Lawes,Chapon,Blake}. 
Another new aspect is the anomalous transport properties associated 
with the onset of helical spin structure in metallic systems such as
$\beta$-MnO$_2$ \cite{Sato}, SrFeO$_3$ \cite{Takeda,Takano}
and MnSi, (Fe,Co)Si \cite{Manyala}. 
These developments urge the microscopic theory of electronic
states to understand the physical properties
associated with the helical spins.

In the absence of the spin-orbit interaction (SOI), one can rotate spin frame 
so that the $z$-axis is parallel to the direction of the local spin.
In this rotated frame, the spins are aligned ferromagnetically
and the original spin structure is reflected in the magnitude and 
phase of the effective transfer integrals. This leads to the 
double exchange interaction \cite{AndersonHasegawa} and various 
phenomena related to the spin chirality \cite{Wen,LeeNagaosa},
respectively. When we consider the state of single 
helical wavevector $q$, the relative angle between the neighboring 
spins does not break the original translational symmetry. 
Furthermore, there is no spin chirality,
i.e., no fictitious magnetic field induced by the solid angle 
subtended by
the spins. Therefore the Hamiltonian in the rotated spin frame 
preserves the periodicity of the original lattice, and hence one 
can define the Bloch wavefunction.

This situation is modified in an essential way when the SOI
is taken into account. In this case,
one cannot rotate the
spin frame with the orbitals being intact, and the transfer integrals
forming a matrix between ions are transformed in a nontrivial way.
Therefore, in general, we expect the incommensurate (IC) 
modulation of the 
transfer integrals and even of the site energies in the effective 
Hamiltonian in the rotated frame.

 Localization/delocalization of electronic state in an IC potential
is an old issue \cite{Sokoloff}. 
Unlike in the case of commensurate periodic potentials, 
the eigen states are not the extended Bloch 
states in the case of IC potentials.
Therefore the band structures would be unusual, i.e., highly 
fragmented, in those IC potentials.
The central issue is whether electronic states are extended or localized
in such kind of potentials, namely, Metal-Insulator Transition (MIT).
Aubry and Andre (A-A) \cite{AuA} have shown that in a simple 1D model 
MIT occurs simultaneously for all energies when the strength of the
IC potential $V_0$ is equal to the transfer integral $t$, i.e., 
if $V_0$ is greater than $t$, the electronic states localize.
We can also regard A-A model as a two dimensional 
tight binding model with IC magnetic flux.
Actually, well-known Hofstadter butterfly is closely related to
this model \cite{Hofstadter}.
Using the {\it trace map} technique,
Kohmoto {\it et al}. \cite{Kohmoto} has exactly studied the scaling properties 
of the Fibonacci lattice system which can be regarded as the A-A model 
with IC modulation $Qa/2\pi$ being the inverse Golden Mean.
Similar problems with IC transfer integral
are also investigated by Kohmoto {\it et al}. \cite{Kohmoto2}.

In this paper, we investigate the localization/delocalization of 
electronic states in IC helical magnets. First we study a model of
5 $d$-orbitals in cubic symmetry taking into account the SOI.
We found that as SOI increases, the localization caused by IC starts 
from the specific $t_{2g}$
wavefunctions at around $q \sim \pi/a$ ($a$: lattice constant). 
In order to scrutinize this localization,
we construct an effective single-band model
for $t_{2g}$ bands.
With this effective model, the localization lengths are studied in more detail
including its dependence on the angle $\varphi$ 
between the spin rotation angle and the helical wavevector. 

\begin{figure}
\includegraphics[width=.8\columnwidth]{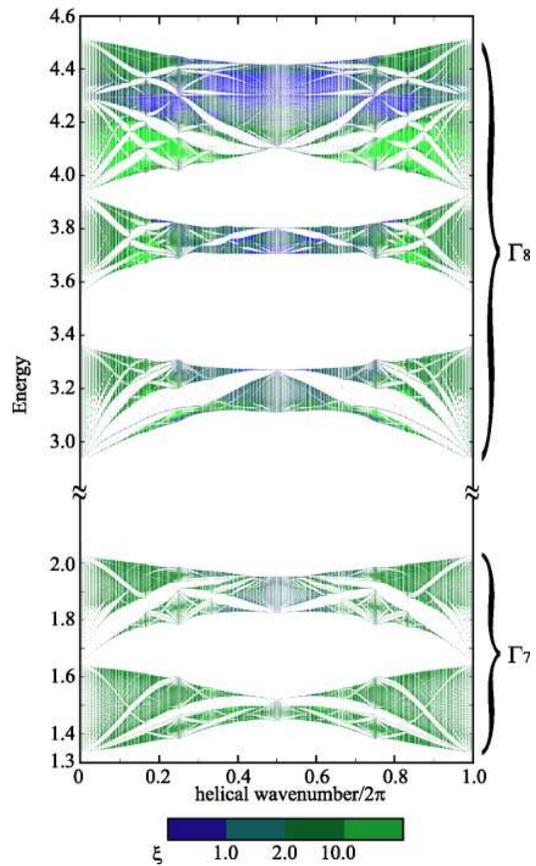}
\caption{Density of states and color map of the localization 
length $\xi$ on it for the d-orbital model Eq. (1).}
\label{ten}
\end{figure}

We start with the following electronic model: 
\begin{eqnarray}
H &=& H_U + H_{SO} + H_d + H_t,
\nonumber \\
H_U &=& -U\sum_j {\vec e_j}\cdot {\vec S_j},
\nonumber \\
H_{SO} &=& -\lambda\sum_j {\vec L_j}\cdot{\vec S_j},
\nonumber \\
H_d &=& \sum_j \epsilon_{\alpha}|d^{\alpha}_{j\sigma}\ra \la d^{\alpha}_{j\sigma}|,
\nonumber \\
H_t &=& t^{\alpha}_{ij}|d^{\alpha}_{j\sigma}\ra \la d^{\alpha}_{j+1,\sigma}|.
\end{eqnarray}
In the octahedral ligand field, the $d$ orbitals are split into 
$e_g$- and $t_{2g}$-orbitals \cite{Kamimura}. 
The $t_{2g}$-orbitals, i.e., $d^{xy},d^{yz}$ and $d^{zx}$,
have energies lower than $e_g$-orbitals, i.e., $d^{x^2-y^2}$, 
and $d^{3z^2-r^2}$ by $10Dq$, 
but the order is reversed as we take the hole picture
in the followings, i.e., $\epsilon_{t_{2g}} - \epsilon_{e_g} = 10 Dq$.
The on-site SOI is considered, 
the matrix elements of
which are calculated by ${\vec L} \cdot {\vec S}$ with $\vec L$ 
($\vec S$)
being the orbital (spin) angular momentum. It is noted that $\vec L$ has
no matrix elements within the $e_g$ sector, while nonzero coupling 
occurs
within $t_{2g}$ sector and between $e_g$ and $t_{2g}$ sectors. 
Considering the hopping between $d$-orbitals and oxygen orbitals \cite{Harrison}, 
we derive the effective transfer integrals $t_{ij}^{\alpha}$
between $d^{\alpha}$-orbitals at neighboring magnetic ions $i$ and $j$. 
We took the values 
$t^{yz}=t^{zx}=0.1$, $t^{3z^2-r^2}=0.3$ and $t^{xy}=t^{x^2-y^2}=0$.
In $H_U$ of Eq. (1), the magnetic moment at site $j$
is described by the unit vector $\vec e_j \equiv (\cos \phi_j \Sin
\theta_j, \Sin \phi_j \Sin \theta_j, \cos \theta_j)$ and $\vec S_j$
denotes the electronic spin operator at site $j$.
We assume the IC helical magnetic structure
for ${\vec S}_j$ along $z$-axis, which is on the spin $(z,x)$-plane,
realized as a result of the frustrated spin exchange interaction.
We focus on the ordered ground state properties, 
and hence the mean field treatment gives a good description of the system.
We assume the ferromagnetic spin configuration perpendicular to the 
helical wavevector $\vec q$, and hence $k_x$, $k_y$ are good quantum numbers,
i.e., the electronic wavefunctions are plane waves along $x$- and 
$y$-directions.
We fix $k_x=k_y=0$ hereafter, and consider the one-dimensional (1D) model 
only along $z$-direction. 
Fig. 1 shows the calculated density of states as a function of 
the helical wavenumber $q$ with the color
specifying the localization length $\xi$. 
We note here that the sample size is a prime number 199, and helical 
wavenumbers $q$'s are taken to be proximate to the IC values. 
All the bands states from the $e_g$-orbitals are extended due to 
the weak SOI, and hence are omitted in Fig. \ref{ten}.
The green region is the extended
states while the blue one is strongly localized within the scale
of lattice constant. We took the values $10Dq = 3$, $U=1.4$, 
and $\lambda=1.0$.

The density of states are understood as follows. 
The largest splitting between $e_g$ and $t_{2g}$ occurs due to 
the ligand field $10Dq$ in Eq. (1). 
Then the $t_{2g}$ bands are further split into bands of $\Gamma_7$ 
and $\Gamma_8$ origin, the latter of which is upper in energy 
since we take the hole picture.
Then both the bands are further split by the spin exchange field $U$. 

By using the iterative method developed by MacKinnon \cite{MacKinnon},
we can calculate the Green's function $G^{(N)}_{1,N} \equiv \la 1|
(E-H)^{-1}|N \ra$, which connects both ends of the long chain.
$G^{(N)}_{1,N}$ is still a $10\times 10$ matrix and
the Lyapnov exponent, i.e., the inverse of the localization length
$\xi$ is obtained as
$\frac{1}{\xi} \equiv -{\rm lim}_{N \to \infty} \frac{1}{2N}
{\rm ln}{\rm Tr}|G^{(N)}_{1,N}|^2.$
The blue color at around $q \sim \pi/a$ means the strong 
localization along the helical axis. 
When we change $\lambda$,
we still observe the localization down to $\lambda \sim 0.2$.
Therefore we conclude that the localization starts in some part of the 
electronic spectrum at around $q = \pi/a$ as one increases the SOI.
The most remarkable point we can grasp from the above figure 
is the energy dependence of localization/delocalization nature, 
namely that there are both localized and extended states at different 
energies for the same $q$.
This is in sharp contrast to the case of A-A model where all
the states are either extended or localized depending only on 
the ratio  $V_0/t$ as mentioned above.

 In order to study this localization in more depth, we now derive the 
effective model for a limiting case, i.e., $10Dq \gg 3\lambda/2 \gg U \gg t$.
Even though this is not nescessarily a suitable limit for
realisitc systems, it clarifies why $\xi$ depends on the orbitals. 
By taking into account the spin degree of freedom, 
there is six-fold degeneracy of the $t_{2g}$ energy levels. 
Because of the on-site SOI, this
degeneracy is lifted and we have two groups of spin-orbit coupled
states, labeled $\Gamma_7$ and $\Gamma_8$ \cite{Kamimura}.
The two-fold degenerate states, $\Gamma_7$, and the four-fold 
degenerate one, $\Gamma_8$, are given by
$
|3^+ \ra = (|d^{xy}_{\ua}\ra + |d^{yz}_{\da}\ra +i
|d^{zx}_{\da}\ra)/\sqrt{3}
$,
$
|3^- \ra = (|d^{xy}_{\da}\ra - |d^{yz}_{\ua}\ra +i
|d^{zx}_{\ua}\ra)/\sqrt{3}
$,
and
$
|1^+\ra = (|d^{yz}_{\ua}\ra +i|d^{zx}_{\ua}\ra)/\sqrt{2}
$,
$
|1^-\ra = (|d^{yz}_{\da}\ra -i|d^{zx}_{\da}\ra)/\sqrt{2}
$,
$
|2^+\ra = (2|d^{xy}_{\ua}\ra -|d^{yz}_{\da}\ra
-i|d^{zx}_{\da}\ra)/\sqrt{6}
$,
$
|2^-\ra = (2|d^{xy}_{\da}\ra +|d^{yz}_{\ua}\ra
-i|d^{zx}_{\ua}\ra)/\sqrt{6}
$,
respectively, where the quantization axis of spin is 
taken to be the $z$ axis. 
Henceforth, we assume that the spin-orbit coupling in our system is
sufficiently large and focus only on the case where the two multiplets,
i.e., $\Gamma_7$ and $\Gamma_8$, do not hybridize with each other.
 
Now, we construct the normalized state $|g_j \ra$ to minimize 
$\la g_j| -U \vec e_j \cdot \vec S_j |g_j \ra$ 
in the Hilbert space spanned by the states in
$\Gamma_7$ or $\Gamma_8$.
The desired states whose spins are parallel to the unit vector 
$\vec e_j$ are explicitly given for $\Gamma_7$ and $\Gamma_8$ by
\begin{equation}
|g^7_j \ra = \Sin \frac{\theta_j}{2}|3^+_j\ra +
e^{i\phi_j}\cos\frac{\theta_j}{2}|3^-_j\ra
\end{equation}
and
\begin{eqnarray}
|g^8_j \ra &=& e^{-i\frac{3}{2}\phi_j}\cos^3 
\frac{\theta_j}{2}|1^+_j\ra
+e^{+i\frac{3}{2}\phi_j}\sin^3 \frac{\theta_j}{2}|1^-_j\ra
\nonumber \\
&-&\sqrt{3} e^{-i\frac{1}{2}\phi_j} \sin \frac{\theta_j}{2} \cos^2
\frac{\theta_j}{2} |2^+_j\ra
\nonumber \\
&-& \sqrt{3} e^{+i\frac{1}{2}\phi_j} \sin^2 \frac{\theta_j}{2} 
\cos \frac{\theta_j}{2} |2^-_j\ra,
\end{eqnarray}
respectively. Here, subscript $j$ denotes the site number and subscripts
7 and 8 correspond to $\Gamma_7$ and $\Gamma_8$ respectively. 
Using these states, we can derive the effective Hamiltonian
$
H = \sum_n T_n c^{\dagger}_n c_{n+1}+h.c. + V_n c^{\dagger}_n c_n,
$
where $c_n/c^{\dagger}_n$ denotes the renormalized annihilation/creation
operator and the effective transfer integral $T_n$ and site energy 
$V_n$ are given as:
\begin{equation}
T_n = \frac{2 t}{3} \Bigr( \sin \frac{\theta_n}{2} \sin
\frac{\theta_{n+1}}{2} +e^{-i\Delta\phi} \cos\frac{\theta_n}{2}
\cos\frac{\theta_{n+1}}{2}
\Bigl),
\nonumber
\end{equation}
$V_n =-{4 t}/{3}$,
where $\Delta \phi = \phi_n-\phi_{n+1}$ for 
$\Gamma_7$, and 
\begin{eqnarray}
T_n &=& t(e^{i\Delta 
\phi/2}\cos{\frac{\theta_n}{2}}\cos{\frac{\theta_{n+1}}{2}}
+e^{-i\Delta \phi/2}\sin{\frac{\theta_n}{2}}\sin{\frac{\theta_{n+1}}{2}})
\nonumber \\
& & \times
(e^{i\Delta \phi}\cos^2{\frac{\theta_n}{2}}\cos^2{\frac{\theta_{n+1}}{2}}
+e^{-i\Delta \phi}\sin^2{\frac{\theta_n}{2}}\sin^2{\frac{\theta_{n+1}}{2}}),
\nonumber
\end{eqnarray}
$V_n = -t(1+\cos^2\theta_n)$.
for $\Gamma_8$.

As for the $\Gamma_7$ case, we can write down $T_n$ as
$
\frac{2 t}{3} e^{i a_{n,n+1}}\cos\frac{\theta_{n,n+1}}{2},
$ 
where $\theta_{n,n+1}$ is the angle between the two spins $\vec S_n$ and
$\vec S_{n+1}$. The phase $a_{n,n+1}$ is the vector potential generated
by the noncollinear spin configuration, but we can eliminate it by
appropriate gauge transformation. 
Then we can conclude that we have no
incommensurability in our 1D $\Gamma_7$ model. 
\begin{figure}[h]
\includegraphics[width=\columnwidth]{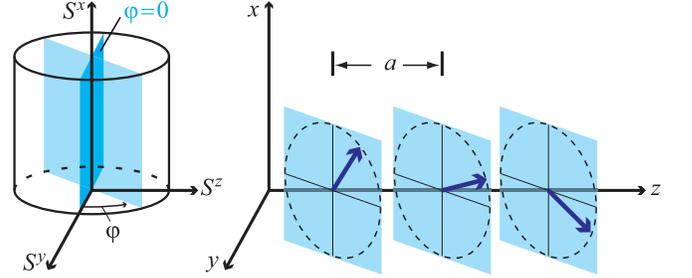}
\caption{Spin plane tilted by angle $\varphi$ from $xy$-plane (left).
The helical spins are rotating on the tilted plane placed periodically 
placed along the $z$-axis.
Blue arrows represent spins, while $a$ denotes the lattice spacing (right).}
\label{co}
\end{figure}
\begin{figure}[h]
\includegraphics[width=\columnwidth,]{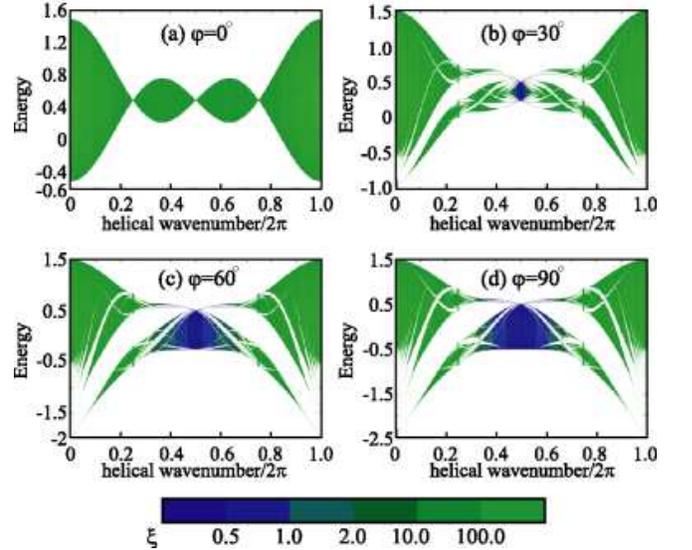}
\caption{Localization length $\xi$ of the effective single band model
for $\Gamma_8$ orbital with (a) $\varphi=0^{\circ}$,
(b) $\varphi=30^{\circ}$,(c) $\varphi=60^{\circ}$, and
(d) $\varphi=90^{\circ}$, respectively.}
\end{figure}

As for the $\Gamma_8$ case, on the other hand, 
the effective site energy $V_n$ explicitly depends on the local 
spin angle $\theta_n$.
If we have the spin configuration in the plane which is 
parallel to the $zx$ plane, i.e., $\theta_n=$const., 
and set the pitch $\Delta
\phi=$const., $V_n$ and $T_n$ are constant. On the other hand, if we
have the tilt of the spin rotation plane from the above plane to the
other plane, $\theta_n$ is no longer a constant and then $V_n$ would
generally be IC. 
$|T_n|$ also depends on both the angles of $\vec S_n$ and $\vec S_{n+1}$.
Here we can conclude that
the case where holes are in $\Gamma_8$, the effective 1D
model would generally be IC.
This explains why the upper part of the $t_{2g}$ density of states in 
Fig. 1 are localized more strongly, where the wavefunction is mainly from 
$\Gamma_8$ components.

Now we focus on the $\Gamma_8$ case and numerically
examine whether the localization of the wavefunction occurs
in more details. We consider the helical spin configuration 
$
\vec S_n = (S \cos(qn),
            S \cos \varphi \hspace{1mm} \sin(qn),
            S \sin \varphi \hspace{1mm} \sin(qn)),
$
where $q$ is helical wavenumber, 
and $\varphi$ denotes the tilt angle of the spin rotating plane
from $xz$-plane (See Fig. \ref{co}). 

The numerical calculations are performed for systems of size
1009, a large prime number, with nearly incommensurate modulations
$q/2\pi=j/1009,\hspace{3mm} (j=1,2,3,...)$. 
The results are shown in Figs. 3, where
the vertical and the horizontal axes represent the energy
and the helical wavenumber, respectively.
We take the unit where $t=1$ and $a=1$. The tilt angles are
$\varphi=0^{\circ},30^{\circ},60^{\circ}$
and $90^{\circ}$ for Figs. 3(a),(b),(c), and (d), respectively.
The energy spectrum in Fig. 3(d) is almost same as 
the lowest band of $\Gamma_8$ bands in Fig. 1.
In Figs. 3, the localization length increases as the
color changes from blue to green. 
The figures clearly display that there are domains 
of strong localization $\xi \sim 1$ when we have a finite 
tilt angle $\varphi$.
On the other hand, for $\varphi=0^{\circ}$ the transfer integrals $T_n$
and the on-site potentials $V_n$ are constants, and there is no
localized states.
Even in the most suitable case for localization,
i.e., Fig.~3(d), however,  
the helical wavenumber $q$ should be
approximately in the range of $2\pi/3 < q < 4 \pi/3$
for the localized states. 
This is because the long period of the helical structure means
the slowly varying and weak perturbations in the rotated frame, and
hence does not cause the localization.

Now we discuss the possible realization of the localized states 
in realistic systems.
From the above results, three important conditions for the localization are 
(i) strong SOI, (ii) short helical period, and (iii) 
the direction of rotating spin plane.
The SOI increases as the mass of the atom get heavier, 
and hence the present model becomes more relevant
from the viewpoint (i). For 3$d$-orbitals of transition metal
oxides, the SOI is typically of the order of 20-30meV, which
is an order of magnitude smaller than the transfer integral $t$.
Therefore the localization length is expected to be rather large, and 
hence the disorder effect such as the impurity scattering might 
hide the IC effect. 
Therefore, even though $\beta$-MnO$_2$
\cite{Sato} and SrFeO$_3$ \cite{Takeda,Takano} show interesting transport
properties, it is unlikely that the localization found in this
paper is relevant to these materials.
4$d$ or 4$f$, 5$f$-orbitals, where SOI is larger than $\sim 0.3$eV, 
are more promising.
Actually there are many rare-earth metals showing the helical 
spin structure such as Tb, Dy, Ho, Er \cite{Koehler,Cowley}.   
From the condition (ii), it is rather hard to find the short period 
helical spin structure. It is typically $4a$-$5a$ or even larger
\cite{Takeda,Manyala,Koehler,Cowley}.
From this viewpoint, MnO$_2$ \cite{Yoshimori,Sato} is an 
interesting case, but the localization is unlikely as discussed
above. As for the condition (iii), we need more study since only
the cubic case has been considered. The directional dependence
of the spin plane might be useful to control the 
localization/delocalization by an external magnetic field. 

Even though the conditions for localization discussed above 
are rather stringent,
which explains why it has never been observed experimentally thus far, 
it will play a vital role 
in the quantum transport properties of the system once realized. 
One direct consequence is the large anisotropy of the 
resistivity between parallel and perpendicular to
the helical axis, i.e., it should be much more
resistive in the parallel direction.

In conclusions, we have studied the localization/delocalization
of the electronic states in helical magnets.
We found the localized states under the condition of
(i) strong spin-orbit interaction, (ii) short helical wavelength,
and (iii) proper direction of the plane on which spins rotate. 
The strong dependence of the localization length $\xi$ 
on the orbital is also found, which is explained by an 
effective model for a certain limiting case.

The authors are grateful to S. Miyashita and K. Azuma for fruitful discussions. 
This work is financially supported by NAREGI Grant, Grant-in-Aids from the
Ministry of Education, Culture, Sports, Science and Technology of
Japan.

\end{document}